# Probing the dynamics of quasicrystal growth using synchrotron live imaging

H. Nguyen-Thi[1], J. Gastaldi[2], T. Schenk[3], G. Reinhart[1], N. Mangelinck-Noel[1], V. Cristiglio[3], B. Billia[1], B. Grushko[4], J. Härtwig[3], H. Klein[5], J. Baruchel[3]

[1]L2MP, UMR CNRS 6137, Université Paul Cézanne Aix-Marseille III, Faculté des Sciences Saint-Jérôme, Case 142, 13397 Marseille Cedex 20, France
[2]CRMCN-CNRS, Campus Luminy Case 913, 13288 Marseille Cedex 9, France
[3]ESRF, BP 220, 38043 Grenoble, France
[4]IFF, Forschungszentrum Juelich GmbH, 52425 Juelich, Germany
[5]Laboratoire de Cristallographie, CNRS Bât. F, BP 166, 38402 Grenoble Cedex 9, France

The dynamics of quasicrystal growth remains an unsolved problem in condensed matter. By means of synchrotron live imaging, facetted growth proceeding by the tangential motion of ledges at the solid-melt interface is clearly evidenced all along the solidification of icosahedral AlPdMn quasicrystals. The effect of interface kinetics is significant so that nucleation and free growth of new facetted grains occur in the melt when the solidification rate is increased. The evolution of these grains is explained in details, which reveals the crucial role of aluminum rejection, both in the poisoning of grain growth and driving fluid flow.

Quasicrystals display long-range orientational order with symmetries (5-fold, 8-fold, …) incompatible with periodicity, previously considered as strictly forbidden. They also exhibit many specific properties (high hardness, low electric conductivity, low friction coefficient …) that have motivated engineering efforts to use them as new materials (e.g. for advanced surface coatings or catalysis). The main challenge of quasicrystal physics is to elucidate how the quasiperiodic order can extend up to the centimetre size of the grains routinely grown nowadays [1-3]. Whether the formation of the stable quasicrystal structure is constrained by local growth rules [4] or by the establishment of long-range atomic correlation [5] is still undetermined. It is therefore critical and timely to deepen the understanding of its dynamical formation during growth from the alloy melt. Because translation symmetry is lacking, it seems unlikely that quasicrystals build up by the attachment of single atoms, thereby generating the whole network just like crystals do. However, as icosahedral clusters have been recently identified in quasicrystal-forming liquids [6], the idea of growth by the attachment of these clusters at the liquid-solid interface [7] is reviving, this process settling the quasicrystalline structure. Besides, the post mortem observation of grains displaying facets perpendicular to the symmetry axes (5, 3, 2 for icosahedral quasicrystals) also suggests that quasicrystals may grow by the forward movement of facets [8,9]. These features were not accessible in the direct observation of rapid solidification at high liquid undercooling [10], as growth was dendritic.

In order to get a clear insight on the shape of the growing grains and the morphology of the solid-liquid interface we carried out the first live probing of the growth of icosahedral AlPdMn quasicrystals by recording synchrotron radiographs during upward Bridgman solidification. In the present paper, the dynamical evolution of the quasicrystal grains is characterized.

Observations were performed by *in situ* and real time synchrotron X-ray radiography at the ID 19 beamline of the European Synchrotron Radiation Facility (ESRF) in Grenoble, France. They consisted in recording all along growth absorption and phase contrast images delivered in monochromatic mode by the highly coherent X-ray beam [11] after crossing a sample solidifying between two graphite foils. The images were recorded either continuously with a CCD camera [12] or at intervals on High Resolution films. The Bridgman directional solidification set-up allowed independent control of both the pulling velocity and the temperature gradient [13]. Two 700 μm-thick sheets were made by grinding rods of $Al_{72.4}Pd_{20.5}Mn_{7.1}$ alloy, whose composition is known to give the icosahedral quasicrystal phase[1]. These sheets were first melted, and then solidified at various pulling velocities (0.4 – 3.6 μm/s) under a temperature gradient of 35 K/cm. Quasicrystallinity was checked by recording Laue patterns and Energy-Dispersive Spectra.

First, in situ X-ray imaging unambiguously proves that the quasicrystals do grow with a facetted solid-melt interface (Fig. 1). In steady-state growth at 0.4 µm/s (Fig. 1a), two grains are growing simultaneously upwards, and the interface between the liquid and these grains shows a cusp at the level of the grain boundary (the radiographs basically show projections along the incident X-ray beam). From the outline of the solid – melt interface, facets and facet edges can be identified on grain 1 and grain 2, which are dodecahedra extending in the direction of pulling (the dodecahedron superposed on Fig. 1a gives the orientation of grain 2). Then, as the applied pulling velocity is increased to 3.6 µm/s, a solidification transient takes place in which the growth velocity (Fig. 2) and grain shape progressively adapt. After about 600 s (Fig. 1b), the solidification front has on the one hand globally receded to a lower temperature, and the growth rate reached the new pulling velocity. On the other hand, the cusp has evolved into a wide and deep liquid groove and while advancing the facets have developed evidences of ledge growth (tangential motion of macrosteps, or bunches of macrosteps), namely the striations running parallel to the solid – liquid interface and the notches neatly visible on the left side of grain 2 (see insert in Fig. 1b, and Fig. 1c). Models based on normal growth with "thick" diffuse interface region [5-7] are not suited to describe ledge growth. The striations and notches are resulting from the repeated kinetic instability of the facets leading to the development of skeletal shapes, as in solution growth [14]. Because of locally higher undercooling and solid – liquid interface roughness, ledges are generated at facet edges and vertices. Then, the tangential motion of these ledges scraps the chemical species rejected in the melt upon solidification, mostly aluminum, down over the facets. This process has a self-poisoning effect on ledge spreading, eventually until cessation. The observed striations thus delineate arrests of ledges, that merely appear as notches when viewed from the side.

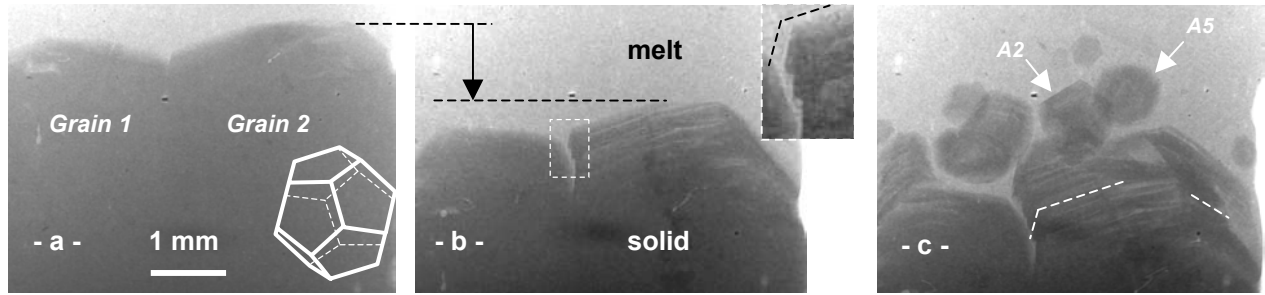

**FIG. 1.** Images recorded by *in situ* and real-time X-ray radiography of AlPdMn quasicrystals growing from their melt : (a) at the end of a growth sequence achieved with a pulling rate V = 0.4 µm/s, (b) and (c) during the growth sequence at V = 3.6 µm/s. The front recoil due to the increase of the applied pulling velocity is shown by the downward arrow in Fig. 1b. The dashed lines in the insert of Fig. 1b and in Fig. 1c are to guide the eyes, and allow orientation by referring to the dodecahedron superposed on Fig. 1a.

Furthermore, the front recoil caused by the increase of the pulling velocity (downward arrow in Fig. 1b), indicates that the attachment of the building elements in the melt to the quasicrystal solidification front is uneasy, and needs significant departure from thermodynamic equilibrium to drive the growth process. As ledge limited growth is characterised by linear kinetics [15], assuming identical solutal recoil [16] the velocity jump $\Delta V$ and temperature shift $\Delta T$ between Fig. 1a and Fig. 1b can be related by the relation $\Delta V = -\mu \Delta T$ with $\mu$ the kinetic coefficient. As far as we are aware, this relation for the first time allows to directly derive a realistic estimate of the kinetic coefficient. Indeed, using $\Delta V = 3.2$ µm/s and $\Delta T = -3.5$ K deduced from the experiment, we get $\mu = 0.9$ µm.s$^{-1}$.K$^{-1}$. This value is two orders of magnitude larger than the value derived by Dong et al. [17] using the Avrami approach of isothermal phase transformation. As this method was applied to the late stage of the transformation, this discrepancy is attributed to the slowing down of the process when grain growth was continuously poisoned by aluminum rejection reducing the effective undercooling, and grain impingement developed, characterized by strong overlapping of the diffusion fields of the various grains. Actually, live synchrotron observation shows that quasicrystal growth kinetics is not so sluggish as admitted but rather comparable to the ledge growth kinetics of semiconductors and oxides ($\mu = 0.826$ µm.s$^{-1}$.K$^{-1}$ for $Bi_4Ge_3O_{12}$ [18]), definitely much slower than the solidification kinetics of pure metals.

Second, 60 sec after Fig. 1b small facetted quasicrystals nucleate and grow freely in the melt just ahead of grains 1 and 2 (Fig. 1c). By their growth these grains provoke solutal blockage due to the mutual impediment of the evacuation of the rejected aluminum, which thus accumulates in the narrow spaces between the grains. Blockage effectively begins with the screening of grains 1 and 2 (Fig. 2), like in the regular columnar-to-equiaxed transition commonly observed in casting [19]. It is worth noticing the black contrast lines decorating the grains in the radiographs (Fig. 3a-c), which are revealing the segregation of the densest and thus most X-ray absorbing component, namely palladium, at grain edges and vertices. Besides, the geometry of the contrast lines enables the determination of grain orientation (Fig. 3d,e), as done for grain 2 in Fig. 1 at an early stage of its development. This method is somewhat similar to the one used to orient dendritic icosahedral AlMn grains from optical metallographs [20]. The new facetted grains are dodecahedral polyhedra often oriented with either a twofold (e.g. A2 in Fig. 1c and Fig. 3b) or fivefold axis (e.g. A5 in Fig. 1c and Fig. 3a) along the incident X-ray beam. A threefold axis is more rarely observed.

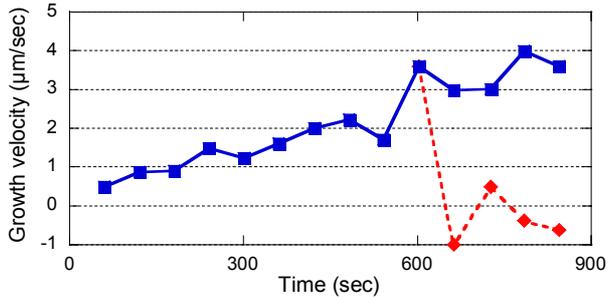

**FIG. 2.** Evolution of the growth velocity of grain 1 in the solidification transient following the abrupt increase of the applied pulling rate from V = 0.4 to V = 3.6 µm/s. The dashed curve (♦) shows the rapid blockage of the growth of the right part of grain 1 by a newly nucleated grain, which is visible in Fig. 1c.

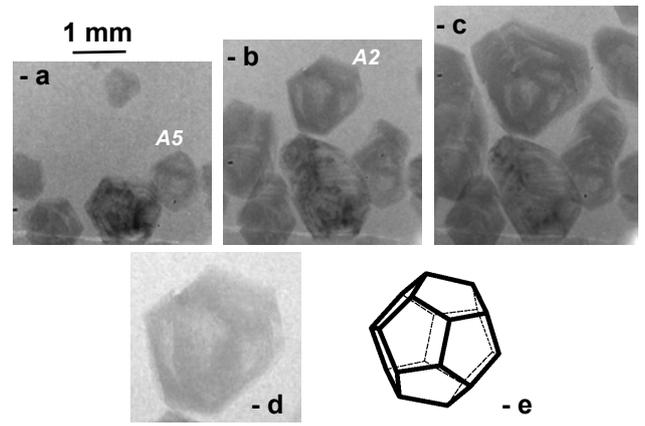

**FIG. 3.** Typical evolution of a free grain after nucleation in the melt, showing progressive blockage by neighbors : – a) 180 sec, - b) 421 sec, - c) 843 sec after nucleation. Pulling rate V = 3.6 µm/s. Grain orientation deduced from radiograhy contrast (d) is shown by projected dodecahedron (e).

The velocity of each facet or edge making the outline of a free grain can be measured along the normal to its trace on the radiographs (Fig. 4). All the facet and edge velocities initially increase with time and ultimately fall down when a facet and a neighbor grain are growing towards each other. Facet growth and grain growth progressively interfere due to the increasing overlap of the diffusion fields, and the facet velocity reaches zero when impingement occurs. At short times, the growth of the new grain is free but not isotropic, which is most obvious for the edges propagating upwards and downwards (arrows 1 and 6 respectively). This asymmetry is the very signature of thermosolutal convection, well documented for dendrites [21]. Indeed, the rejection of both aluminum and latent heat renders the melt surrounding the solid grain lighter, thus creating a driving force for growth-induced natural convection. For the downward - growing edge, this buoyancy - driven convection sweeps the surrounding fluid away upwards, and brings much undercooled melt in contact with this edge, which causes it to grow faster. For the upward-growing edge, the low-undercooling fluid enveloping the grain flows upwards into the path of the edge which causes it to propagate more slowly. The dodecahedron shape is preserved as long as grain growth is free, i.e. not suffering neighbor interaction altering the contour. Indeed, in the process of grain impingement, the facets facing each other undergo progressive smoothening that gives way to curved boundaries. Meanwhile, the facets seeing open melt ahead, that generally grow at a small angle to the pulling direction, conversely persist and continue to advance. This stage, which may even reach a steady-state as indicated by the plateaus in the velocities of facet 1 and facet 2, results in grain elongation (Fig. 3b,c).

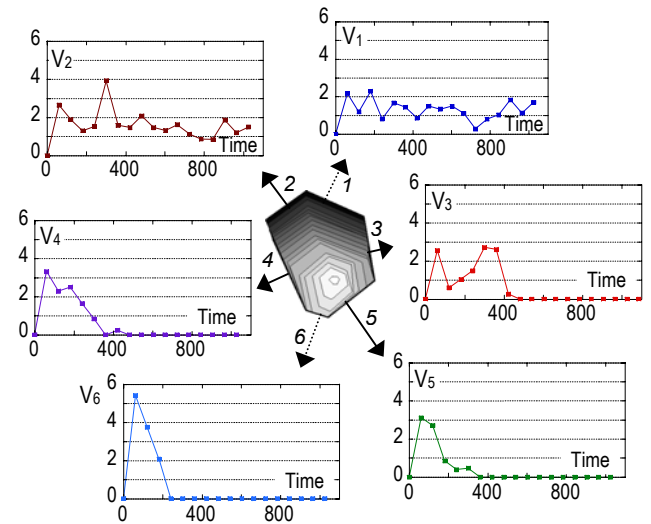

**FIG. 4.** Variation with time (in sec) of the velocity (in µm/sec) of the facets (arrows 2 – 5, velocities $V_2 – V_5$) and edges (arrows 1 and 6, velocities $V_1$ and $V_6$) making the outline of the highest grain in Fig. 3a. The origin of time is set at grain nucleation, 540 s after applying the pulling rate V = 3.6 µm/s.

The mere fact that the nucleation of new grains does not occur during the growth sequence at V = 0.4 µm/s but only after the increase to V = 3.6 µm/s brings out fundamental information. Namely, it is because the kinetic undercooling needed for quasicrystal growth has exceeded the critical nucleation undercooling that the birth of new grains is enabled in the melt just above grains 1 and 2. Due to the topological similarity between the local structural order of icosahedral AlPdMn and the icosahedral clusters in the melt [6,22], a very low resistance of icosahedral quasicrystals to nucleation is expected. This suggests that, aside the possibility of heterogeneous nucleation on the graphite foils that cannot be readily discarded, the new grains might grow from such clusters so that the nature of quasicrystal nucleation, which remains presently uncertain, may be peculiar : heterogeneous because on solid embryos, homogeneous because these embryos are clusters structurally self-embedded in the melt and not added refining particles. However, a critical nucleation radius comparable to the icosahedral cluster size implies an extremely small solid – liquid interface energy, so that larger clusters [6,23] formed in the melt by individual clusters sharing atoms or glued together seem better candidates for the nucleation embryos.

Using synchrotron live imaging to probe in situ and in real-time the dynamics of icosahedral AlPdMn quasicrystal growth from the melt, the facetted character of grain growth, which results from ledge spreading, is established. From the solid – melt interface undercooling, a realistic estimate of the kinetic coefficient is deduced. Rather than by local heat flow as it is largely believed, the growth of quasicrystals is controlled by interface kinetics and aluminum diffusion, as further evidenced in the growth of new AlPdMn quasicrystals.

The authors express their gratitude to T. Bactivelane, R. Chagnon for technical assistance in the experiments. They also thank A.A. Chernov, M. Kléman, V. Dmitrienko and M. de Boissieu for fruitful discussions.